\documentclass[fleqn,twoside]{article}
\usepackage[headings]{espcrc2}
\usepackage{cite}

\readRCS
$Id: espcrc2.tex,v 1.2 2004/02/24 11:22:11 spepping Exp $
\usepackage{graphicx}
\usepackage[figuresright]{rotating}

\newcommand{\AmS}{{\protect\the\textfont2
  A\kern-.1667em\lower.5ex\hbox{M}\kern-.125emS}}

\title{Status of jet cross sections to NNLO}

\author{Stefan Weinzierl\address{Institut f{\"u}r Physik, Universit{\"a}t Mainz,
        D - 55099 Mainz, Germany}
       }
       
\runtitle{Jet cross sections to NNLO}
\runauthor{S. Weinzierl}

\begin{document}

\begin{abstract}
I review the state-of-the-art for fully differential numerical NNLO programs.
Topics which are covered include the calculation of two-loop amplitudes,
multiple polylogarithms, cancellation of infra-red divergences at NNLO and the
efficient generation of the phase space.
Numerical results for $e^+ e^- \rightarrow \;\mbox{2 jets}$ are also discussed.
\vspace{1pc}
\end{abstract}

% typeset front matter (including abstract)
\maketitle

\section{INTRODUCTION}

In the near future
the LHC experiments will provide us with a large sample of multi-particle
final states.
In order to extract information from this data, precise theoretical calculations
are necessary.
This implies to extend perturbative calculations for selected processes
from next-to-leading order (NLO)
to next-to-next-to-leading order (NNLO) in the perturbative expansion in the strong
coupling constant.
Due to a large variety of interesting jet observables it is desirable not to perform
this calculation for a specific observable, but to set up a computer
program, which yields predictions for any infra-red safe observable relevant to the process
under consideration.

The past years have witnessed a tremendous progress in techniques 
for the computation of two-loop integrals
and in the calculation of two-loop amplitudes 
\cite{Dobbs:2004bu}.
In addition, several options for the cancellation of infra-red divergences have been discussed
\cite{Kosower:2002su,Kosower:2003cz,Weinzierl:2003fx,Weinzierl:2003ra,Anastasiou:2003gr,Gehrmann-DeRidder:2005cm,Binoth:2004jv,Heinrich:2006sw,Kilgore:2004ty,Frixione:2004is,Somogyi:2005xz}.
Among those, the subtraction method -- well-known from NLO computations
\cite{Frixione:1996ms,Catani:1997vz,Dittmaier:1999mb,Phaf:2001gc,Catani:2002hc} --
and sector decomposition 
\cite{Binoth:2000ps}
are the most promising candidates.
First numerical results have become available
for the processes $e^+ e^- \rightarrow \mbox{2 jets}$ and $H \rightarrow \gamma \gamma$
\cite{Anastasiou:2004qd,Anastasiou:2004xq,Anastasiou:2005qj,Weinzierl:2006ij}.

\section{TWO-LOOP AMPLITUDES}

The master formula to calculate an observable at an collider with no 
initial-state hadrons (e.g. an electron-positron collider) is given by
\begin{eqnarray}
\label{master_formula}
\lefteqn{
\langle {\cal O}^{(j)} \rangle = \frac{1}{2 K(s)}
             \frac{1}{\left( 2 J_1+1 \right)}
             \frac{1}{\left( 2 J_2+1 \right)}
} & & \nonumber \\
& &
             \sum\limits_n
             \int d\phi_{n-2}
             {\cal O}^{(j)}_n\left(p_1,...,p_n\right)
             \sum\limits_{helicity} 
             \left| {\cal A}_n \right|^2
\end{eqnarray}
where $p_1$ and $p_2$ are the momenta of the initial-state particles, 
$2K(s)=2s$ is the flux factor and $s=(p_1+p_2)^2$ is the center-of-mass energy squared.
The factors $1/(2J_1+1)$ and $1/(2J_2+1)$ correspond
to an averaging over the initial helicities. $d\phi_{n-2}$ is the invariant phase space measure
for $(n-2)$ final state particles and
${\cal O}^{(j)}_n\left(p_1,...,p_n\right)$ is the observable, evaluated with an $n$-parton configuration.
The index $j$ indicates that the leading order contribution depends on $j$ partons.
The observable has to be infra-red safe, in particular this implies that 
we must have
in single unresolved limits 
\begin{eqnarray}
{\cal O}^{(j)}_{n+1}(p_1,...,p_{n+1}) & \rightarrow & {\cal O}^{(j)}_n(p_1',...,p_n'),
\end{eqnarray}
and in double unresolved limits
\begin{eqnarray}
{\cal O}^{(j)}_{n+2}(p_1,...,p_{n+2}) & \rightarrow & {\cal O}^{(j)}_n(p_1',...,p_n').
\end{eqnarray}
${\cal A}_n$ is the amplitude with $n$ partons.
At NNLO we need the following expansions of the amplitudes:
\begin{eqnarray}
& & 
\hspace*{-6mm}
 \left| {\cal A}_n \right|^2
  =  
   \left. {\cal A}_n^{(0)} \right.^\ast {\cal A}_n^{(0)} 
 + 
   \left(
             \left. {\cal A}_n^{(0)} \right.^\ast {\cal A}_n^{(1)} 
           + \left. {\cal A}_n^{(1)} \right.^\ast {\cal A}_n^{(0)} 
       \right)
 \nonumber \\
 & &
 + 
   \left(
             \left. {\cal A}_n^{(0)} \right.^\ast {\cal A}_n^{(2)} 
           + \left. {\cal A}_n^{(2)} \right.^\ast {\cal A}_n^{(0)}  
           + \left. {\cal A}_n^{(1)} \right.^\ast {\cal A}_n^{(1)} 
       \right),
 \nonumber \\
& &
\hspace*{-6mm}
  \left| {\cal A}_{n+1} \right|^2
 =  
   \left. {\cal A}_{n+1}^{(0)} \right.^\ast {\cal A}_{n+1}^{(0)}
 \nonumber \\
 & &
 + 
   \left(
          \left. {\cal A}_{n+1}^{(0)} \right.^\ast {\cal A}_{n+1}^{(1)} 
        + \left. {\cal A}_{n+1}^{(1)} \right.^\ast {\cal A}_{n+1}^{(0)}
 \right),
 \nonumber \\ 
& &
\hspace*{-6mm}
  \left| {\cal A}_{n+2} \right|^2
 = 
   \left. {\cal A}_{n+2}^{(0)} \right.^\ast {\cal A}_{n+2}^{(0)}. 
\end{eqnarray}
Here ${\cal A}_n^{(l)}$ denotes an amplitude with $n$ partons and $l$ loops.
The master formula eq.~(\ref{master_formula}) is rewritten
symbolically as 
\begin{eqnarray}
\langle {\cal O}^{(j)} \rangle & = & 
             \sum\limits_n \int {\cal O}^{(j)}_{n} \; d\sigma_{n}
\end{eqnarray}
and the LO, NLO and NNLO contribution as 
\begin{eqnarray}
\label{def_LO_NLO_NNLO}
 & & 
 \hspace*{-5mm}
\langle {\cal O}^{(j)} \rangle^{LO} = \int {\cal O}^{(j)}_{n} \; d\sigma_{n}^{(0)},
 \nonumber \\
 & & 
 \hspace*{-5mm}
\langle {\cal O}^{(j)} \rangle^{NLO} = \int {\cal O}^{(j)}_{n+1} \; d\sigma_{n+1}^{(0)} + \int {\cal O}^{(j)}_{n} \; d\sigma_{n}^{(1)},
 \nonumber \\
 & & 
 \hspace*{-5mm}
\langle {\cal O}^{(j)} \rangle^{NNLO} = \int {\cal O}^{(j)}_{n+2} \; d\sigma_{n+2}^{(0)} 
 \nonumber \\
 & &
                   + \int {\cal O}^{(j)}_{n+1} \; d\sigma_{n+1}^{(1)} 
                   + \int {\cal O}^{(j)}_{n} \; d\sigma_{n}^{(2)}.
\end{eqnarray}
To compute the two-loop amplitudes, new techniques have been developed.
Examples include the application of the Mellin-Barnes transformation \cite{Smirnov:1999gc,Bierenbaum:2003ud},
differential equations \cite{Gehrmann:1999as},
nested sums\cite{Moch:2001zr,Weinzierl:2002hv,Weinzierl:2004bn,Moch:2005uc},
or sector decomposition \cite{Binoth:2000ps}.
In addition, several methods, which reduce the work-load have been invented.
Prominent examples are
integration-by-parts-identities\cite{Chetyrkin:1981qh,Tkachov:1981wb},
reduction algorithms \cite{Tarasov:1996br,Laporta:2001dd}
or the cut technique \cite{Bern:2000dn}.
In the results, like for example in the two-loop amplitude for
$e^+ e^- \rightarrow \;\mbox{3 jets}$ \cite{Garland:2002ak,Moch:2002hm}, a new class of mathematical functions appears.
These are the multiple polylogarithms.

\section{MULTIPLE POLYLOGARITHMS}

The multiple polylogarithms are defined by
\begin{eqnarray} 
\label{multipolylog2}
\lefteqn{
 \mbox{Li}_{m_1,...,m_k}(x_1,...,x_k)
  = } & & \nonumber \\
 & &
 \sum\limits_{i_1>i_2>\ldots>i_k>0}
     \frac{x_1^{i_1}}{{i_1}^{m_1}}\ldots \frac{x_k^{i_k}}{{i_k}^{m_k}}.
\end{eqnarray}
Alternatively, they can be defined by an integral representation. We introduce the auxiliary functions
\begin{eqnarray}
\label{Gfuncdef}
 \lefteqn{
G(z_1,...,z_k;y) = } & & \nonumber \\
 & &
 \int\limits_0^y \frac{dt_1}{t_1-z_1}
 \int\limits_0^{t_1} \frac{dt_2}{t_2-z_2} ...
 \int\limits_0^{t_{k-1}} \frac{dt_k}{t_k-z_k}.
\end{eqnarray}
In this definition 
one variable is redundant due to the following scaling relation:
\begin{eqnarray}
G(z_1,...,z_k;y) & = & G(x z_1, ..., x z_k; x y)
\end{eqnarray}
With the short-hand notation
\begin{eqnarray}
\label{Gshorthand}
 \lefteqn{
G_{m_1,...,m_k}(z_1,...,z_k;y)
 = } & & \nonumber \\
 & &
 G(\underbrace{0,...,0}_{m_1-1},z_1,...,z_{k-1},\underbrace{0...,0}_{m_k-1},z_k;y)
\end{eqnarray}
One then finds
\begin{eqnarray}
\label{Gintrepdef}
 \lefteqn{
\mbox{Li}_{m_1,...,m_k}(x_1,...,x_k)
 = } & & \\
 & &
 (-1)^k 
 G_{m_1,...,m_k}\left( \frac{1}{x_1}, \frac{1}{x_1 x_2}, ..., \frac{1}{x_1...x_k};1 \right). 
 \nonumber 
\end{eqnarray}
Multiple polylogarithms have interesting mathematical properties. 
In particular, there are two Hopf algebra structures defined on them\cite{Moch:2002hm,Weinzierl:2003jx,Weinzierl:2006qs}.
At the end of the day of an analytic calculation physicists would like to get a number.
This requires a method for the numerical evaluation of multiple polylogarithms.
The simplest example is
the numerical evaluation of the dilogarithm \cite{'tHooft:1979xw}:
\begin{eqnarray}
\mbox{Li}_{2}(x) & = & - \int\limits_{0}^{x} dt \frac{\ln(1-t)}{t}
 = \sum\limits_{n=1}^{\infty} \frac{x^{n}}{n^{2}}
\end{eqnarray}
The power series expansion can be evaluated numerically, provided $|x| < 1.$
Using the functional equations 
\begin{eqnarray}
 & & \hspace*{-5mm}
\mbox{Li}_2(x) =  -\mbox{Li}_2\left(\frac{1}{x}\right) -\frac{\pi^2}{6} -\frac{1}{2} \left( \ln(-x) \right)^2,
 \\
 & & \hspace*{-5mm}
\mbox{Li}_2(x) = -\mbox{Li}_2(1-x) + \frac{\pi^2}{6} -\ln(x) \ln(1-x), \nonumber 
\end{eqnarray}
any argument of the dilogarithm can be mapped into the region
$|x| \le 1$ and
$-1 \leq \mbox{Re}(x) \leq 1/2$.
The numerical computation can be accelerated  by using an expansion in $[-\ln(1-x)]$ and the
Bernoulli numbers $B_i$:
\begin{eqnarray}
\mbox{Li}_2(x) & = & \sum\limits_{i=0}^\infty \frac{B_i}{(i+1)!} \left( - \ln(1-x) \right)^{i+1}.
\end{eqnarray}
The generalisation to multiple polylogarithms proceeds along the same lines \cite{Vollinga:2004sn}:
Using the integral representation of
\begin{eqnarray}
G_{m_1,...,m_k}\left(z_1,z_2,...,z_k;y\right)
\end{eqnarray}
one
transforms all arguments into a region, where one has a converging power series expansion.
To accelerate the convergence of the series the H\"older convolution is used\cite{Borwein}.

\section{INFRA-RED DIVERGENCES}

The individual contributions on the r.h.s. of eq.~(\ref{def_LO_NLO_NNLO})
to $\langle {\cal O}^{(j)} \rangle^{NLO}$ and $\langle {\cal O}^{(j)} \rangle^{NNLO}$ are in general infra-red divergent, only the sum is finite.
However, these contributions live on different phase spaces, which prevents a naive Monte Carlo approach.
To render the individual contributions finite, one adds and subtracts suitable chosen terms.
The NLO contribution is given by
\begin{eqnarray}
 & & 
 \hspace*{-5mm}
\langle {\cal O}^{(j)} \rangle^{NLO} = 
   \int \left( {\cal O}^{(j)}_{n+1} \; d\sigma_{n+1}^{(0)} - {\cal O}^{(j)}_{n} \circ d\alpha^{(0,1)}_{n} \right)
 \nonumber \\
  & &
 + \int \left( {\cal O}^{(j)}_{n} \; d\sigma_{n}^{(1)} + {\cal O}^{(j)}_{n} \circ d\alpha^{(0,1)}_{n} \right).
\end{eqnarray}
The notation ${\cal O}^{(j)}_{n} \circ d\alpha^{(0,1)}_{n}$ is a reminder, that
in general the approximation is a sum of terms
\begin{eqnarray}
{\cal O}^{(j)}_{n} \circ d\alpha^{(0,1)}_{n} & = & \sum {\cal O}^{(j)}_{n} \; d\alpha^{(0,1)}_{n}
\end{eqnarray}
and the mapping used to relate the $n+1$ parton configuration to a $n$ parton configuration
differs in general for each summand.
In a similar way, the NNLO contribution is written as
\begin{eqnarray}
\langle {\cal O}^{(j)} \rangle^{NNLO} & = &
 \langle {\cal O}^{(j)} \rangle^{NNLO}_{n+2}
+  \langle {\cal O}^{(j)} \rangle^{NNLO}_{n+1}
 \nonumber \\
 & &
+  \langle {\cal O}^{(j)} \rangle^{NNLO}_{n},
\end{eqnarray}
with
\begin{eqnarray}
\lefteqn{
 \langle {\cal O}^{(j)} \rangle^{NNLO}_{n+2}
 =  
 \int \left( {\cal O}^{(j)}_{n+2} \; d\sigma_{n+2}^{(0)} 
             - {\cal O}^{(j)}_{n+1} \circ d\alpha^{(0,1)}_{n+1}
 \right. } & & \nonumber \\
 & &
 \left.
             - {\cal O}^{(j)}_{n} \circ d\alpha^{(0,2)}_{n} \right),
 \nonumber \\
 \lefteqn{
 \langle {\cal O}^{(j)} \rangle^{NNLO}_{n+1}
 =  
 \int \left( {\cal O}^{(j)}_{n+1} \; d\sigma_{n+1}^{(1)} 
               + {\cal O}^{(j)}_{n+1} \circ d\alpha^{(0,1)}_{n+1}
 \right. } & & \nonumber \\
 & & \left.
               - {\cal O}^{(j)}_{n} \circ d\alpha^{(1,1)}_{n} \right),
 \nonumber \\
 \lefteqn{
 \langle {\cal O}^{(j)} \rangle^{NNLO}_{n}
 = 
 \int \left( {\cal O}^{(j)}_{n} \; d\sigma_n^{(2)} 
               + {\cal O}^{(j)}_{n} \circ d\alpha^{(0,2)}_{n}
 \right. } & & \nonumber \\
 & & \left.
               + {\cal O}^{(j)}_{n} \circ d\alpha^{(1,1)}_{n} \right).
\end{eqnarray}
$d\alpha^{(0,1)}_{n+1}$ is the NLO subtraction term for $(n+1)$-parton configurations,
$d\alpha^{(0,2)}_{n}$ and $d\alpha^{(1,1)}_{n}$ are generic NNLO subtraction terms.
It is convenient to split these terms into
\begin{eqnarray}
 d\alpha^{(0,2)}_{n} & = & d\alpha^{(0,2)}_{(0,0),n} - d\alpha^{(0,2)}_{(0,1),n},
 \nonumber \\
 d\alpha^{(1,1)}_{n} & = & d\alpha^{(1,1)}_{(1,0),n} + d\alpha^{(1,1)}_{(0,1),n},
\end{eqnarray}
such that $d\alpha^{(0,2)}_{(0,0),n}$ and $d\alpha^{(1,1)}_{(1,0),n}$ approximate 
$d\sigma_{n+2}^{(0)}$ and $d\sigma_{n+1}^{(1)}$, respectively.
$d\alpha^{(0,2)}_{(0,1),n}$ and $d\alpha^{(1,1)}_{(0,1),n}$ are approximations to
$d\alpha^{(0,1)}_{n+1}$.

\section{PHASE SPACE GENERATION}

It is a well-known fact, that in the collinear limit spin correlations remain.
For example, the spin-dependent splitting functions for $g \rightarrow q \bar{q}$ reads
\begin{eqnarray}
 P^{(0,1)}_{g \rightarrow q \bar{q}} & = & 
   \frac{2}{s_{ij}} \left[ -g^{\mu\nu} + 4 z (1-z) \frac{k^\mu_\perp k^\nu_\perp}{k_\perp^2} \right].
\end{eqnarray}
The collinear limits occurs for $k_\perp^2 \rightarrow 0$.
The term
\begin{eqnarray}
 {\cal A}_\mu \frac{1}{s_{ij}} \frac{k_\perp^\mu k_\perp^\nu}{k_\perp^2} {\cal A}_\nu
\end{eqnarray}
is proportional to the spin correlation.
In four dimensions the spin-averaged splitting functions are obtained by integrating over the azimuthal angle $\varphi$ 
of $p_i$ around $p$.
By using spin-averaged antenna functions, the subtraction terms have not the same point-wise singular behaviour as the matrix
elements, which is required for  local subtraction terms.
Instead, cancellations of singularities occurs only after an integration over the azimuthal angle 
over all collinear splittings 
of the matrix elements.
For $n$ final-state particles, this is a one-dimensional integration in the $(3n-4)$-dimensional phase space.
It can be shown, that in the single collinear limit, the spin correlation depends on the azimuthal angle $\varphi$
as
\begin{eqnarray}
 {\cal A}_\mu \frac{1}{s_{ij}} \frac{k_\perp^\mu k_\perp^\nu}{k_\perp^2} {\cal A}_\nu
 & \sim &
 C_0 + C_2 \cos( 2 \varphi + \alpha).
\end{eqnarray}
One can therefore perform the average with two points, where the azimuthal angle takes the values
$\varphi$ and $\varphi + \pi/2$,
while all other coordinates remain fixed.
In detail this is done as follows:
We partition the phase space into different channels.
Within one channel, the phase space is generated iteratively according to
\begin{eqnarray}
 d\phi_{n+1} & = & d\phi_n d\phi_{Dipole\;i,j,k}.
\end{eqnarray}
For each channel we require that the product $s_{ij} s_{jk}$ is the smallest among all considered channels and that
$s_{ij} < s_{jk}$. Therefore it follows that with channel $(i,j,k)$ also channel $(k,j,i)$ has to be included into
the partioning of the phase space.
For the dipole phase space measure we have
\begin{eqnarray}
 d\phi_{dipole}
 = \frac{s_{ijk}}{32 \pi^3} 
       \int\limits_0^1 dy \; \left(1-y\right)
       \int\limits_0^1 dz \; 
       \int\limits_0^{2\pi} d\varphi.
\end{eqnarray}
We can therefore generate the $(n+1)$-parton configuration from the $n$-parton configuration by using three random numbers
$u_1$, $u_2$, $u_3$ and by setting
\begin{eqnarray}
 y = u_1, \;\;\; z = u_2 \;\;\; \varphi = 2 \pi u_3.
\end{eqnarray}
This defines the invariants as
\begin{eqnarray}
& &
 s_{ij} = y s_{ijk},
 \;\;\;
 s_{ik} = z (1-y) s_{ijk}, 
 \nonumber \\
 & &
 s_{jk} = (1-z) (1-y) s_{ijk}.
\end{eqnarray}
From these invariants and the value of $\varphi$ we can reconstruct the four-momenta of the $(n+1)$-parton configuration
\cite{Weinzierl:1999yf}.
The additional phase space weight due to the insertion of the $(n+1)$-th particle is
\begin{eqnarray}
 w & = & \frac{s_{ijk}}{16 \pi^2} \left( 1-y \right).
\end{eqnarray}
We have therefore a parameterization of the phase space, such that for every collinear limit the azimuthal average can
be easily performed, while keeping all other coordinates fixed.
It is clear that this procedure can be iterated for multiple collinear emissions.

\section{NUMERICAL RESULTS}

A test case for the methods discussed above for the cancellation of infra-red divergences and numerical
phase space integration is the process $e^+ e^- \rightarrow \;\mbox{2 jets}$.
The two-jet cross section
has the perturbative expansion
\begin{eqnarray}
 \lefteqn{
 \langle \sigma \rangle^{(2-jet)}
 = } & & \\
 & &
 \langle \sigma \rangle^{(0)} \left( 1 + \frac{\alpha_s}{2\pi} B^{(2-jet)}
                               + \left( \frac{\alpha_s}{2\pi} \right)^2 C^{(2-jet)} \right).
 \nonumber 
\end{eqnarray}
$\langle \sigma \rangle^{(0)}$ is the total hadronic cross section at leading order, the numerical value
at LEP energies $\sqrt{Q^2} = 91.187$ GeV is
$\langle \sigma \rangle^{(0)} = 40807.4\;\mbox{pb}$.
The jets are defined
according to the Durham jet algorithm with $y=0.01$.
For the numerical values of 
the NLO and NNLO coefficients $B^{(2-jet)}$ and $C^{(2-jet)}$ one obtains
\begin{eqnarray}
 B^{(2-jet)} & = & -13.674 \pm 0.004,
 \nonumber \\
 C^{(2-jet)} & = & -231.6 \pm 0.3.
\end{eqnarray}
The correctness of this result can be verified with the help of the known results for the 
total hadronic cross section at NNLO,
the three-jet cross section at NLO and the four-jet cross section at LO. 
We have the perturbative expansions
\begin{eqnarray}
\lefteqn{
 \langle \sigma \rangle^{(tot)}
 =} & & \nonumber \\
 & &
 \langle \sigma \rangle^{(0)} \left( 1 + \frac{\alpha_s}{2\pi} B^{(tot)}
                               + \left( \frac{\alpha_s}{2\pi} \right)^2 C^{(tot)} \right),
 \nonumber \\
 \lefteqn{
 \langle \sigma \rangle^{(3-jet)}
 = } & & \nonumber \\
 & &
 \langle \sigma \rangle^{(0)} \left( \frac{\alpha_s}{2\pi} B^{(3-jet)}
                               + \left( \frac{\alpha_s}{2\pi} \right)^2 C^{(3-jet)} \right),
 \nonumber \\
 \lefteqn{
 \langle \sigma \rangle^{(4-jet)}
 = \langle \sigma \rangle^{(0)} \left( \frac{\alpha_s}{2\pi} \right)^2 C^{(4-jet)}.} & &
\end{eqnarray}
The values of the coefficients are:
\begin{eqnarray}
 & &
\hspace*{-7mm}
 B^{(tot)} = 2,
 \;
 C^{(tot)} = 5.64,
 \\
 & &
\hspace*{-7mm}
 B^{(3-jet)} = 15.679 \pm 0.004,
 \nonumber \\
 & &
\hspace*{-7mm}
 C^{(3-jet)} = 153.2 \pm 0.4,
 \;
 C^{(4-jet)} = 84.39 \pm 0.05. 
 \nonumber 
\end{eqnarray}
We must have
\begin{eqnarray}
 B^{(tot)} & = & B^{(2-jet)} + B^{(3-jet)},
 \nonumber \\
 C^{(tot)} & = & C^{(2-jet)} + C^{(3-jet)} + C^{(4-jet)}.
\end{eqnarray}
We find
\begin{eqnarray}
 & & \hspace*{-5mm}
 B^{(2-jet)} + B^{(3-jet)} = 2.005 \pm 0.006,
 \nonumber \\
 & & \hspace*{-5mm}
 C^{(2-jet)} + C^{(3-jet)} + C^{(4-jet)}  = 6.0 \pm 0.5,
\end{eqnarray}
which agrees well with the values of $B^{(tot)}$ and $C^{(tot)}$.

\section{OUTLOOK}

Fully differential numerical programs at NNLO are a challenging task.
I reviewed the state of the art and discussed the calculation of two-loop amplitudes,
the cancellation of infra-red divergences and methods for an efficient generation of the phase space.
As a first application I presented complete results for the NNLO correction to the process
$e^+ e^- \rightarrow \;\mbox{2 jets}$, based on the subtraction method.
This gives us confidence, that the extension to other processes
like $e^+ e^- \rightarrow \mbox{3 jets}$ or 
$p p \rightarrow \mbox{2 jets}$ is within reach.

% references

\end{document}